\documentclass[conference]{IEEEtran}
\IEEEoverridecommandlockouts
\usepackage{cite}
\usepackage{enumitem}
\usepackage{amsmath,amssymb,amsfonts}
\usepackage{algorithmic}
\usepackage{algorithm,algorithmic,refcount}

\usepackage{graphicx}
\usepackage{textcomp}
\usepackage{color, soul}
\usepackage[dvipsnames,table,xcdraw]{xcolor}
\usepackage{subfigure}
\usepackage{siunitx}
\usepackage[nocomma]{optidef}
\def\BibTeX{{\rm B\kern-.05em{\sc i\kern-.025em b}\kern-.08em
    T\kern-.1667em\lower.7ex\hbox{E}\kern-.125emX}}

\newcommand\blfootnote[1]{%
 \begingroup
 \renewcommand\thefootnote{}\footnote{#1}%
 \addtocounter{footnote}{-1}%
  \endgroup
}

 \makeatletter
\def\footnoterule{\kern-3\p@
  \hrule \@width 2in \kern 2.6\p@} 
\makeatother

\begin{document}

\title{Behind Closed Doors: Process-Level Rootkit Attacks in Cyber-Physical Microgrid Systems}

\author{
\IEEEauthorblockN{\textbf{Suman Rath}\IEEEauthorrefmark{1}, \textbf{Ioannis Zografopoulos}\IEEEauthorrefmark{2}, \textbf{Pedro P. Vergara}\IEEEauthorrefmark{3},\\ \textbf{Vassilis C. Nikolaidis}\IEEEauthorrefmark{4}, \textbf{Charalambos Konstantinou}\IEEEauthorrefmark{2}}
\IEEEauthorblockA{
\IEEEauthorblockA{\IEEEauthorrefmark{1}School of Electrical \& Computer Engineering, Oklahoma State University\\
\IEEEauthorrefmark{2}CEMSE Division, King Abdullah University of Science and Technology (KAUST)\\
\IEEEauthorrefmark{3}Intelligent Electrical Power Grids, Delft University of Technology (TU Delft)\\
\IEEEauthorrefmark{4}Dept. of Electrical and Computer Engineering, Democritus University of Thrace}
E-mail: suman.rath@okstate.edu, p.p.vergarabarrios@tudelft.nl, vnikolai@ee.duth.gr\\ \{ioannis.zografopoulos, charalambos.konstantinou\}@kaust.edu.sa}
}

\IEEEaftertitletext{\vspace{-1.1\baselineskip}}
\maketitle

\begin{abstract}
Embedded controllers, sensors, actuators, advanced metering infrastructure, etc. are cornerstone components of cyber-physical energy systems such as microgrids (MGs). Harnessing their monitoring and control functionalities, sophisticated schemes enhancing MG stability can be deployed. However, the deployment of `smart' assets increases the threat surface. Power systems possess mechanisms capable of detecting abnormal operations. Furthermore, the lack of sophistication in attack strategies can render them detectable since they blindly violate power system semantics. On the other hand, the recent increase of process-aware rootkits that can attain persistence and compromise operations in undetectable ways requires special attention. In this work, we investigate the steps followed by stealthy rootkits at the process level of control systems pre- and post-compromise. We investigate the rootkits' precompromise stage involving the deployment to multiple system locations and aggregation of system-specific information to build a neural network-based virtual data-driven model (VDDM) of the system. Then, during the weaponization phase, we demonstrate how the VDDM measurement predictions are paramount, first to orchestrate crippling attacks from multiple system standpoints, maximizing the impact, and second, impede detection blinding system operator situational awareness. 

\blfootnote{Accepted paper at IEEE Power \& Energy Society General Meeting 2022}

\end{abstract}


\vspace{-2mm}
\section{Introduction}
\vspace{-1mm}
Microgrids (MGs), among others, foster the inclusion of renewable energy sources. Closely coupled cyber and physical layers guarantee the increased flexibility and robust operation of MGs. 
The information and communication cyber layer receives inputs from sensors and issues controls in the physical layer. \textcolor{black}{As a result, adversaries can exploit cyber vulnerabilities (e.g., insecure protocols, software bugs, etc.) to port their attacks impacting MG's control and stability  \cite{rath2020cyber}.  Attacks able to manipulate sensor data can have significant impacts (e.g., blackouts, human safety, equipment damage, etc.) \cite{konstantinou2017gps}.}

\textcolor{black}{Existing literature focuses on various cyberattacks, their exploitation techniques, and schemes to detect and mitigate them \cite{zografopoulos2021cyber}. The severity and risks associated with malware infections within industrial systems, especially rootkits, has captured the attention of researchers \cite{powers2015whitelist}. Rootkits represent a class of malware which can intelligently hide their presence inside their targets \cite{8988665}. They can eavesdrop system data and allow attackers to collect real-time system information via remotely accessible connections \cite{xenofontos2021consumer}. Adversaries can exploit these connections to issue malicious commands, manipulate the infected device(s), and stealthily control their operation.}

\textcolor{black}{MG-based process level rootkit attacks should be considered when designing and implementing detection and mitigation strategies. 
Rootkit attacks should represent a crucial part of reliability assessment procedures for certification laboratories, tasked with identifying and eliminating vulnerabilities in embedded control devices and designing experimental testbeds for evaluation of cyber-physical power systems\cite{powers2015whitelist, reeves2012intrusion}.} This paper extends our previous work-in-progress on stealthy rootkit attacks \cite{10.1145/3447555.3466576}, and highlights a potential rootkit attack path after the installation of the malware at multiple locations within the MG. After its deployment, the rootkit aggregates system measurements to build an accurate system replica allowing the estimation of the MG states trajectories; in our case, using a neural network-based approach. We demonstrate different approaches that rootkits could utilize to conceal their presence (during the system information aggregation phase) and disguise the attack impact overcoming existing security fortifications exploiting system state estimations. The paper presents simulation results to demonstrate the attack methodology and achievable system impact.
\vspace{-2mm}
\section{Attack Model}
\vspace{-2mm}

\begin{figure}
\centering
  \includegraphics[width=0.7\linewidth]{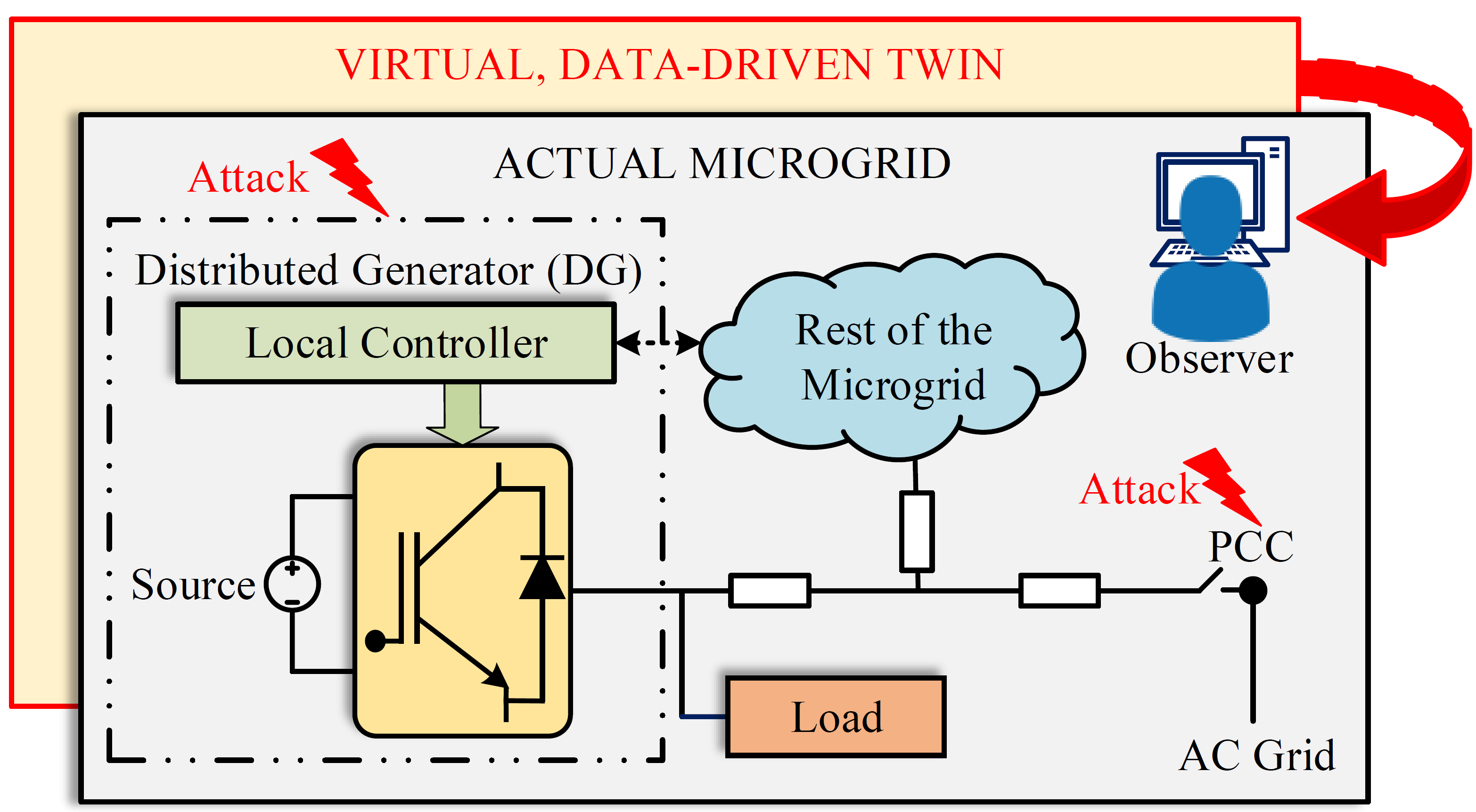}
  \vspace{-2mm}
  \caption{{Microgrid (MG) rootkit attack model.}}
  \vspace{-5mm}
  \label{fig:teaser1}
\end{figure}

\textcolor{black}{Rootkits can leverage the vulnerabilities (cyber layer) of operating system architectures 
encountered in power system workstations, HMIs, etc. during their deployment. Persistence is then achieved by masking their presence through system information and log modifications \cite{MITRE}. In most cases, the compromised systems exhibit vulnerabilities similar to consumer computers, and rootkits can exploit them to be deployed on different levels of the hardware or software stack, e.g., firmware, bootloader, memory, kernel -based rootkits, etc. In Fig. \ref{fig:teaser1}, we present a MG model infected with the rootkit. Our threat model assumes that the rootkit can access and stealthily modify process level sensor measurements and controller strategies at different locations within the MG, i.e., a strong adversary model \cite{zografopoulos2021cyber}. The rootkit achieves persistence and disguises its operation, by collecting sufficient system data to anticipate the MG state trajectories. By reporting the expected values, the rootkit remains undetected by system operators, while maximizing its potential attack impact.}

In this work, we build on the concept of process level rootkit attacks hiding their presence in cyber-physical MGs \cite{10.1145/3447555.3466576}. Once the rootkit gains access to various MG elements (e.g., controllers, actuators, etc.), it starts eavesdropping on the system information. Such process level data could include state parameters (e.g., voltage, frequency, etc.) utilized to build a virtual data-driven model (VDDM), similar to a digital twin of the entire MG. The granularity of the VDDM depends on the attacker's goal, capabilities, and available resources. The model is then employed to predict future system states and control operations. The attacker uses the predicted values of the VDDM MG to bypass security mechanisms and remain undetected. The attack aims to create gradual changes at the local distributed generation (DG) level whose effects can propagate through grid devices (e.g., controllers, inverters, etc.) and services impacting the overall power system. 

The changes can range from minor alterations, like voltage injections with minor effects on system performance, to noise additions in sensor measurements which can  generate erroneous control inputs to grid devices affecting the power generation, voltage stability, and introducing harmonic frequencies \cite{kuruvila2020hardware}. Although such perturbations might be small in magnitude to remain undetected, their impact can increase over prolonged periods of time and threaten nominal system operation \cite{9595243}.

The malware is able to intelligently coordinate the manipulation of power electronic converters at both the DG level and the point of common coupling (PCC) to maximize its impact. For example, the manipulation of sensor measurements at the PCC level can be used to trigger false alarms (e.g., indicating a fault condition) and force the MG to operate in islanded/autonomous mode (i.e., by tripping its circuit breaker). However, in such operation mode the MG is more vulnerable since it will not be able to synchronize its frequency and voltage setpoints using grid values as references. Thus, adversaries can compromise local controllers and sensors to create voltage and frequency instabilities. 
Additionally, manipulation of load sharing patterns among different power generation resources within the MG can disrupt the optimal scheduling \cite{lakshminarayana2022loadaltering}. In Algorithm \ref{a:alg}, we present the  post-installation attack methodology followed by the rootkit. 

The rootkit operates non-intrusively inside the host until the assignment of a malicious target objective. For instance, if the rootkit aims to create frequency deviations, it will use the MG's VDDM to identify the subset of agents that will achieve this objective. It will then manipulate power and frequency sensors providing inputs to the DG-level inverter control. The attack targeting multiple DGs, continues until the desired level of instability is achieved. After the target objective is completed, the rootkit modifies the measurements before being reported to system observers (to disguise its presence) and remains inactive until a new attack is performed.

\setlength{\textfloatsep}{0pt}
\begin{algorithm}[t!]
\caption{Process level rootkit attack methodology.}
	\begin{algorithmic}[1]
		\STATE Eavesdrop vulnerable devices to collect measurements. 
		\STATE Use acquired measurements to construct a VDDM of the system. Train ANN-based estimator for state prediction.
		\STATE Identify malicious objective (e.g., frequency/voltage instability, disturbances in optimal load sharing, etc.).
		\STATE Use the VDDM to identify the devices which can be manipulated to achieve the set objective.
		\STATE Determine a time frame for the attack 
		when the system is vulnerable and/or rootkit actions will not trigger alarms. 
		\STATE Modify sensor measurements at the PCC level to create artificial fault conditions, forcing defensive islanding.
		\STATE During autonomous operation, alter the frequency and voltage
		references and the setpoints of power devices.
		\STATE Modify sensor measurements and use the ANN-based state estimator to predict the nominal system state.
		\STATE Mask rootkit manipulation by reporting the \textit{predicted} values to the MG observer(s). 
		
	\end{algorithmic}
	\label{a:alg}
	\vspace{-1mm}
\end{algorithm}

\vspace{-2mm}
\section{Control and Attack Formulation} \label{s:controlArch}
\vspace{-2mm}

The control framework of the MG has a hybrid structure with a central master controller at the PCC-level and one distributed local controller at each DG. This structure consists of primary, secondary, and tertiary control mechanisms. The primary controller is responsible for the load sharing among the  DGs, and has a droop-based control objective which is achieved through  frequency and voltage regulation as follows:
\vspace{-4mm}
\begin{align}
\vspace{-1mm}
\omega^{\ast}_i = \omega_n-D_{P_i}P_i\\
v^{\ast}_{i} = v_{n}-D_{Q_i}Q_i
\vspace{-1mm}
\end{align}
where \(\omega^{\ast}_i\) and \(v^{\ast}_{i}\) represent frequency and voltage of \(i^{th}\) DG, \(\omega_n\) and \(v_{n}\) denote nominal frequency and voltage, and \(P_i\) and \(Q_i\) the active and reactive power measurements of the \(i^{th}\) DG. \(D_{P_i}\) and \(D_{Q_i}\) are the droop coefficients  determined considering the following (for a \(N\)-DG MG):
\begin{align}
\vspace{-1mm}
    D_{P_1}P_1 = D_{P_2}P_2 = ... = D_{P_N}P_N = \Delta\omega_{Th}\\
    D_{Q_1}Q_1 = D_{Q_2}Q_2 = ... = D_{Q_N}Q_N = \Delta{v}_{Th}
    \vspace{-1mm}
\end{align}
where \(\Delta\omega_{Th}\) and \(\Delta{v}_{Th}\) represent the maximum permissible deviation of frequency and voltage, respectively.

In the grid-connected mode, the grid determines frequency and voltage values \cite{bidram2017cooperative}. In this case, the load sharing pattern is regulated through the modification of \(\omega_n\) and \(v_{n}\). As demonstrated in Eqs. (1) and (2), the actions of the primary controller will cause a droop in the trajectory of frequency and voltage as the values of \(P\) and \(Q\) increase. To solve this issue and restore the system parameters to their nominal trajectories, the secondary controller introduces \(\delta\omega\) and \(\delta{v}\) terms in the power controller equations: 
\begin{align}
\omega^{\ast}_i = \omega_n-D_{P_i}P_i + \delta\omega \\
v^{\ast}_{i} = v_{n}-D_{Q_i}Q_i + \delta{v}
\end{align}
In Eqs. (5) and (6), \(\delta\omega\) and \(\delta{v}\) negate the droop caused as a result of the primary control. They are defined as follows \cite{rath2020cyber}:
\begin{equation} \label{eq1}
\begin{split}
\delta\dot{\omega} &= K_1\Big(\sum_{j\epsilon{N(i)}}{a_{ij}}(\omega_{j}-\omega_i)+g_i(\omega_n-\omega_i) \\
 & + \sum_{j\epsilon{N(i)}}{a_{ij}}(D_{P_j}P_{j}-D_{P_i}P_{i})\Big)
\end{split}
\end{equation}
\begin{align}
\delta\dot{v} = K_2\Big(\sum_{j\epsilon{N(i)}}{a_{ij}}(D_{Q_j}Q_{j}-D_{Q_i}Q_{i})\Big)
\end{align}
\textcolor{black}{where \({a_{ij}}\) represents the element 
in the adjacency matrix of the bidirectional connected communication graph}, \(g\) represents the pinning gain, and \(K_1\), \(K_2\) are constants \cite{rath2020cyber}.
The control targets for the secondary controller are shown below:
\begin{align}
\lim_{t \to \infty}||\omega_i(t)-\omega_n|| = 0 \;\forall\; i \\
\lim_{t \to \infty}||D_{Pi}{P_i}-D_{Pj}{P_j}|| = 0 \;\forall\; i,\;j \\
\lim_{t \to \infty}||D_{Qi}{Q_i}-D_{Qj}{Q_j}|| = 0 \;\forall\; i,\;j
\end{align}

The rootkit's long-term target is to disrupt the aforementioned control objectives by introducing small manipulations whose consequences are unnoticeable by bad data detectors (BDDs). 
The rootkit methodology and the steps required for the attack are discussed below.

\vspace{-1mm}
\subsection{VDDM: Target Identification and State Prediction}
\vspace{-1mm}

In recent days, virtual power plants (VPPs) have been widely used for the modeling of demand-response management schemes, participation in day-ahead markets, etc. \cite{mnatsakanyan2014novel, baringo2018day}. Open source data enable adversaries to access and build such VPPs for malicious gains \cite{keliris2019open}. In our use case, the adversary uses the VDDM of the MG for two purposes: \emph{(i)} performance evaluation of an attack vector before its introduction in the physical system, and \emph{(ii)} prediction of system behaviour under normal conditions to hide the rootkit's manipulations from the system, e.g., BDDs.

Before initiating an attack, the adversary attempts to infect the maximum possible agents at the sensor, communication, and controller levels to maximize possible data collection. However, in a practical scenario, the attacker's ability to install malware at different stages may be limited by physical, hardware, or software constraints \cite{wang2016malicious, anubi2019enhanced}. In that case, the attacker can construct a virtual replica of the system with limited state information. Thus, the virtual system may not be a precise replica of the actual system and its accuracy will be determined by the quality of the data captured and the adversary's knowledge about the system \cite{zografopoulos2021cyber}. The replica, in our work, uses a combination of Kalman filtering (KF) and neural networks to predict the future states of the MG. Specifically, KF is used to train an artificial neural network (ANN) that acts as the state estimator. \textcolor{black}{It improves the quality and size of the training dataset by estimating missing values of data \cite{habtom1997estimation}. KF can also generate a sufficiently large dataset from a relatively smaller set of accurate data points, eliminating any potential uncertainty associated with the rootkit's data collection abilities (which may affect training of the VDDM).}
\textcolor{black}{The hybrid estimator -- combining KF and ANN -- improves prediction accuracy and convergence speed when compared with traditional KF/NN approaches} \textcolor{black}{\cite{sieberg2021hybrid},} \cite{zhan2006neural}. 
The steps involved in the generation of the training and test datasets for the ANN are described below.

Since the attacker has 
access to sensors and  controllers, real-time measurements from a set of infected physical sensors \(\zeta\) can be  captured. The state of the system is estimated from the captured 
measurements using KF 
Let the vector containing the measurements (obtained from \(\zeta\)) be 
\(m\). The system state is: 
\begin{align}
    x(t) = F(t)x(t-1) + B(t)c(t) + n(t)
\end{align}
where \(x(t)\), \(F(t)\), \(B(t)\), \(c(t)\), \(n(t)\) represent the state vector, the state transition matrix, the control input matrix, the control input vector, and the unknown process noise vector corresponding to \(x(t)\), respectively. \(n(t)\) is given by a zero mean normal distribution provided by the covariance matrix \(C_0(t)\), $E[n(t)n(t)'] = C_0(t)$. 
The equation depicting the measurement can be written as follows.
\begin{align}
    m(t) = H(t)x(t) + v(t)
\end{align}
where \(H(t)\) and \(v(t)\) represent the transformation matrix mapping states to measurements and the known process noise vector, respectively. \(v(t)\) is obtained from the covariance matrix \(C_1(t)\), $E[v(t)v(t)'] = C_1(t)$. 
The state in the next time step can be estimated as follows:
\begin{align}
    x(t+1 \vert {t}) = F(t)x(t \vert {t}) + \textcolor{black}{B(t)c(t)} \\
    m(t+1 \vert {t}) = H(t)x(t+1 \vert {t}) \\
    n(t+1) = m(t+1) - m(t+1 \vert {t})
\end{align}
Further,
\begin{align}
    x(t+1 \vert {t+1}) = x(t+1 \vert {t}) + N(t+1)n(t+1)
\end{align}
where \(N(t+1)\) represents the Kalman gain utilized for state covariance. The state covariance can be estimated as: 
\begin{align}
    P(t+1 \vert {t}) &= F(t)P(t \vert {t})F(t)' + C_0(t)\\
    S(t+1) &= H(t+1)P(t+1 \vert {t})H(t+1)' + C_1(t+1)\\
    N(t+1) &= P(t+1)H(t+1)'S(t+1)-1 \\
    P(t+1 \vert {t+1}) &= P(t+1 \vert {t}) - N(t+1)S(t+1)N(t+1)'
\end{align}
where \(P(t+1 \vert {t})\), \(S(t+1)\), \(N(t+1)\) and \(P(t+1 \vert {t+1})\) are the state prediction covariance, measurement covariance, filter gain, and updated state covariance, respectively.

The measurements collected by the rootkit, are used to analyze the current state and estimate the future states of the system. Furthermore, the obtained data is used to perform supervised training of an ANN with a \(l\)-layer feed-forward architecture that can predict the future state \(X_P\) of the system once it receives the 
measurement vector \(m\) and the requisite estimation time \(T\) as inputs. Let the initial set of weights (defined during the training phase) be \(W_0\). The initial output \(X_{P0}\) can be defined as:
\begin{align}
    X_{P0} = f_A(W_0, l, m, T)
\end{align}
As the training progresses, the weights are updated through back-propagation (post-completion of each iteration) to minimize the error in the outputs per the predefined loss function.
\begin{align}
W_{i+1} = f_B(\eta, e, W_i, l)
\end{align}
where \(\eta\) is the learning rate, and \(e\) represents the error between output of the network (after the \(i_{th}\) iteration) and the estimated output from the training data.

The predicted state (output) after the \(i_{th}\) iteration can be defined as a function mapping the obtained measurements (input) to the future state, given a set of weights \(W_i\).
\begin{align}
    X_{Pi} = f_A(W_i, l, m, T)
\end{align}

After convergence, the network can be deployed for state estimation in the malicious system. 
Before the attack vector \(\alpha\) introduction,
the attacker tries to estimate the future system state
\(X_{P_{\alpha},T_{\alpha}}\) using the malicious VDDM. 
\begin{align}
X_{P_{\alpha},T_{\alpha}} = f_A(W_f, l, m + \alpha, T_\alpha)
\end{align}
where
\(T_\alpha\) represents the end of the attack period.
The magnitude of \(\alpha\) must satisfy the following constraint: $|\alpha| \leq |\alpha_{max}|$, where \(|\alpha_{max}|\) represents the maximum allowable magnitude of the attack vector to evade BDDs.
During the weaponization phase, the attacker estimates and reports the expected normal system state 
to conceal the rootkit actions.
\begin{align}
X_{N\alpha,T\alpha} = f_A(W_f, l, m, T_\alpha)
\end{align}
where 
\(X_{N\alpha,T\alpha}\) is the predicted normal trajectory of the system in the absence of the attack vector \(\alpha\).




\vspace{-1mm}
\section{Results and Discussion}

\begin{figure}[t]
\centering
  \includegraphics[width=0.9\linewidth]{
  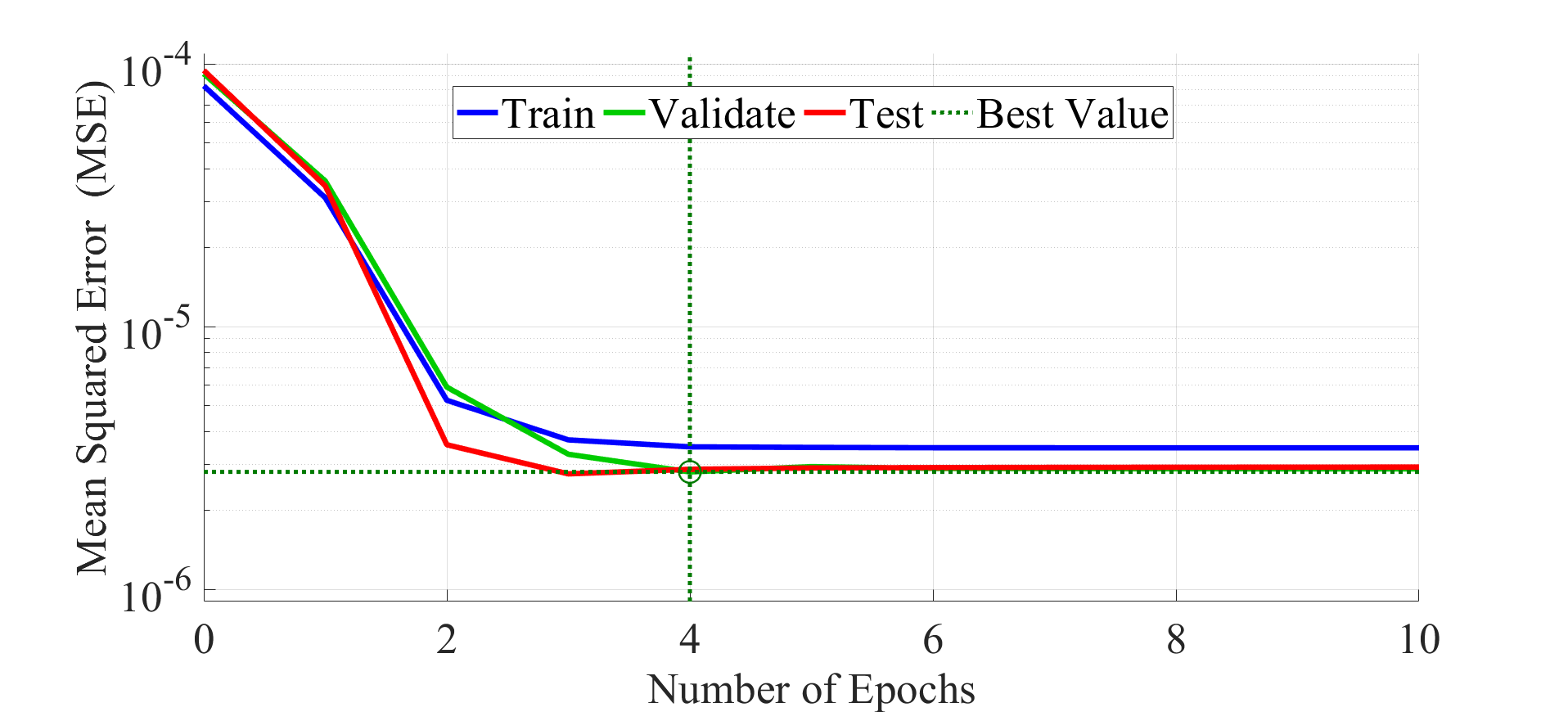}
  \vspace{-2mm}
  \caption{{Mean squared error (MSE) during training, validation and testing of the artificial neural network (ANN).}}
  \label{fig:2}
\end{figure}

An AC MG model is designed using MATLAB following the architecture of the model presented in \cite{rath2020cyber}. The model consists of four DGs. Details of various system parameters can be obtained from \cite{rath2020cyber}. Data collected from this test model is used to generate a VDDM, which can predict the future states of the system using the KF-ANN hybrid state estimator model discussed in Section \ref{s:controlArch}. The ANN is trained through supervised learning using labelled datasets generated from the designed MG model and the KF algorithm. Fig. \ref{fig:2} shows the accuracy of the  VDDM by depicting the mean squared error (MSE) during the training, validation and testing phases.

The MG model is also used as a test system to demonstrate the possible impact that a rootkit malware with partial/complete access can cause to disrupt nominal behavior. Three target objectives are assigned to the malicious rootkit: \emph{(i)} frequency manipulation, \emph{(ii)} voltage manipulation, and \emph{(iii)} disruption of load sharing. For these objectives, the rootkit uses its controller- and sensor-level access at the PCC side to first create a false fault alarm, cause defensive islanding of the system, and then compromise the DG operation.

\vspace{-1mm}
\subsection{Target: Frequency Manipulation }

\begin{figure}[t]
\centering
  \includegraphics[width=0.9\linewidth]{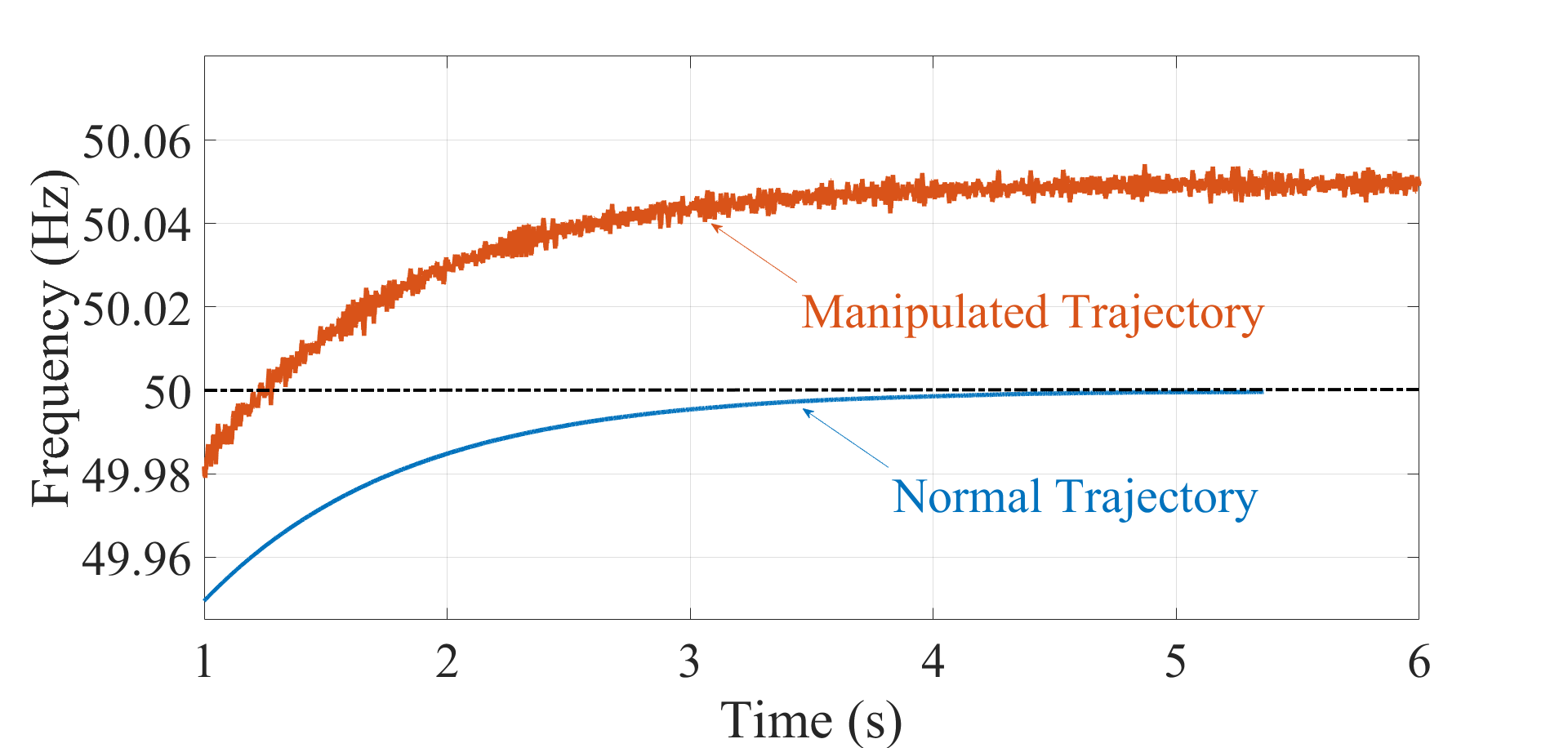}
  \vspace{-2mm}
  \caption{Rootkit attack impact on the DG frequency.} 
  \label{fig:3}
\end{figure}

\textcolor{black}{When the adversary instructs the rootkit to initiate the frequency manipulation attack, the malware evaluates its access level and uses the malicious VDDM to identify the vulnerable agents, which can be exploited to achieve the desired level of deviation. In this case, the adversarial objective entails achieving a higher steady state frequency. To accomplish this, the rootkit manipulates load devices and sensors. It modifies \(\omega_n\) affecting \(\delta\omega\) as per Eq. (7) and further influences \(\omega^{\ast}_i\) as per Eq. (5). Fig. \ref{fig:3} depicts the normal and manipulated system trajectories after forced islanding. 
The rootkit can successfully distort the system frequency trajectory through the manipulation of load sensors and the local controller associated with the leader, i.e., DG 1. Fig. \ref{fig:3} shows that the frequency attains steady state at 50.05 Hz, however, the deviation could be further increased depending on the introduced attack vector.}

\subsection{Target: Voltage Manipulation}

\begin{figure}[t]
\centering
  \includegraphics[width=0.9\linewidth]{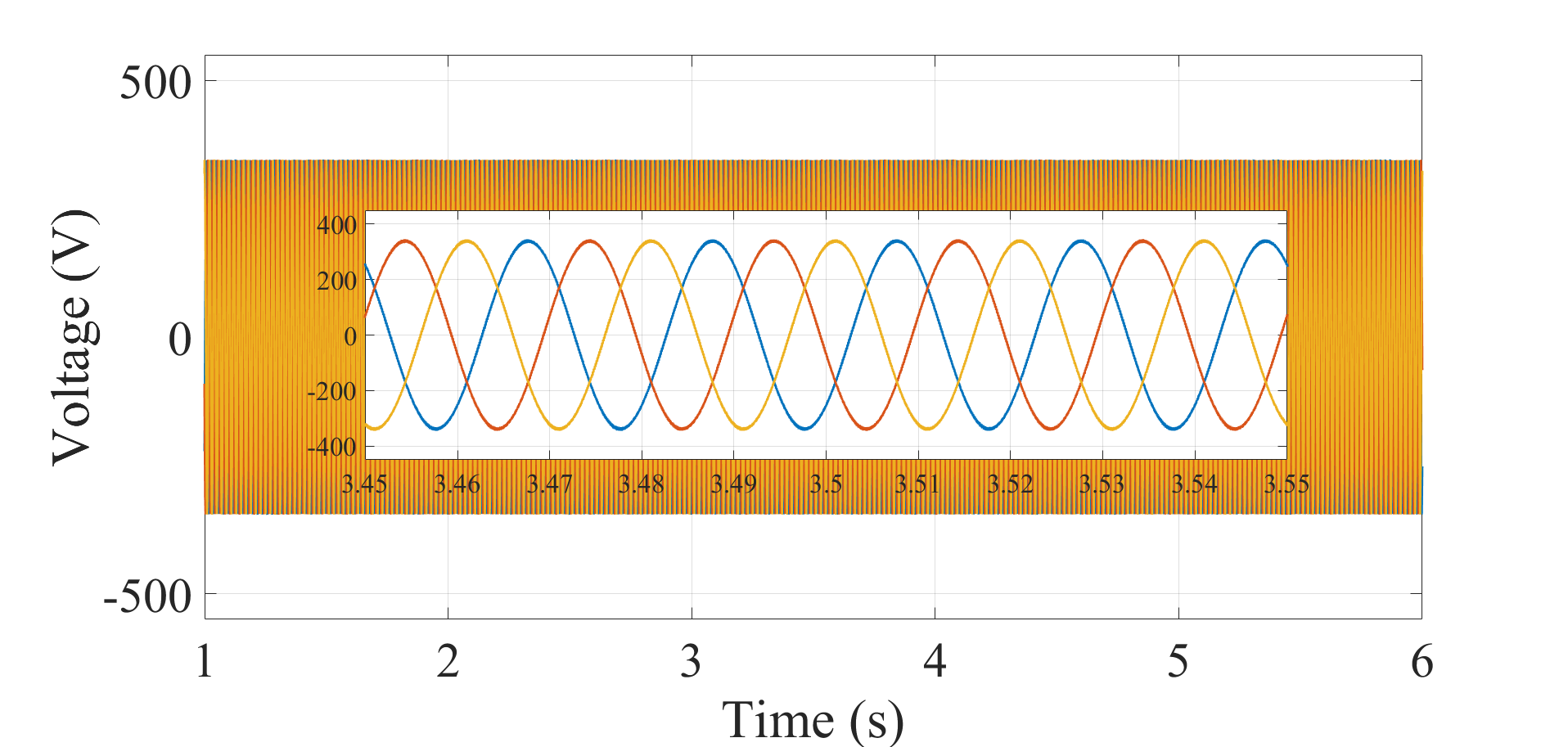}
  \vspace{-4mm}
  \caption{Three-phase voltage magnitude at DG 1 after islanding occurs.}
  \label{fig:4}
\end{figure}

\begin{figure}[t]
\centering
  \includegraphics[width=0.9\linewidth]{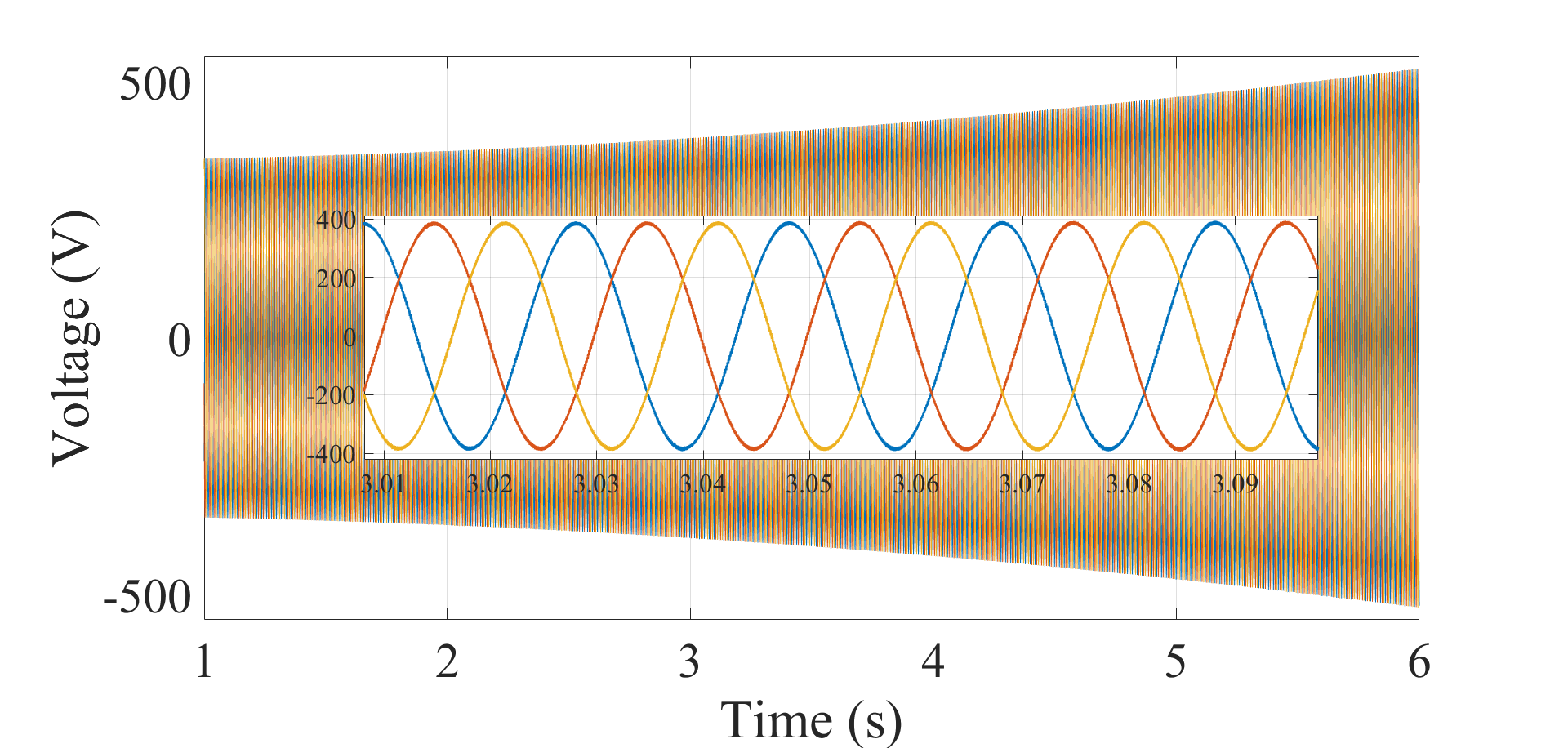}
  \vspace{-4mm}
  \caption{Abnormal voltage increase at DG 1 after the rootkit attack.}
  \label{fig:5}
\end{figure}

In the voltage manipulation case, the rootkit malware can either modify the voltage levels at any particular bus or create voltage instability affecting the whole MG system. After the target assignment, the malware uses the VDDM to determine the manipulation strategy required to achieve the set objective. In our case, the rootkit introduces false reactive power demand by thoroughly crafting bias injections to the system sensors. Fig. \ref{fig:4} and Fig. \ref{fig:5} show the normal voltage trajectory and the voltage trajectory after the rootkit manipulations are initiated. The experimental results demonstrate that the rootkit attack can cause the gradual increase of the three-phase voltage magnitude at DG 1.

\begin{figure}[t]
\centering
  \includegraphics[width=0.9\linewidth]{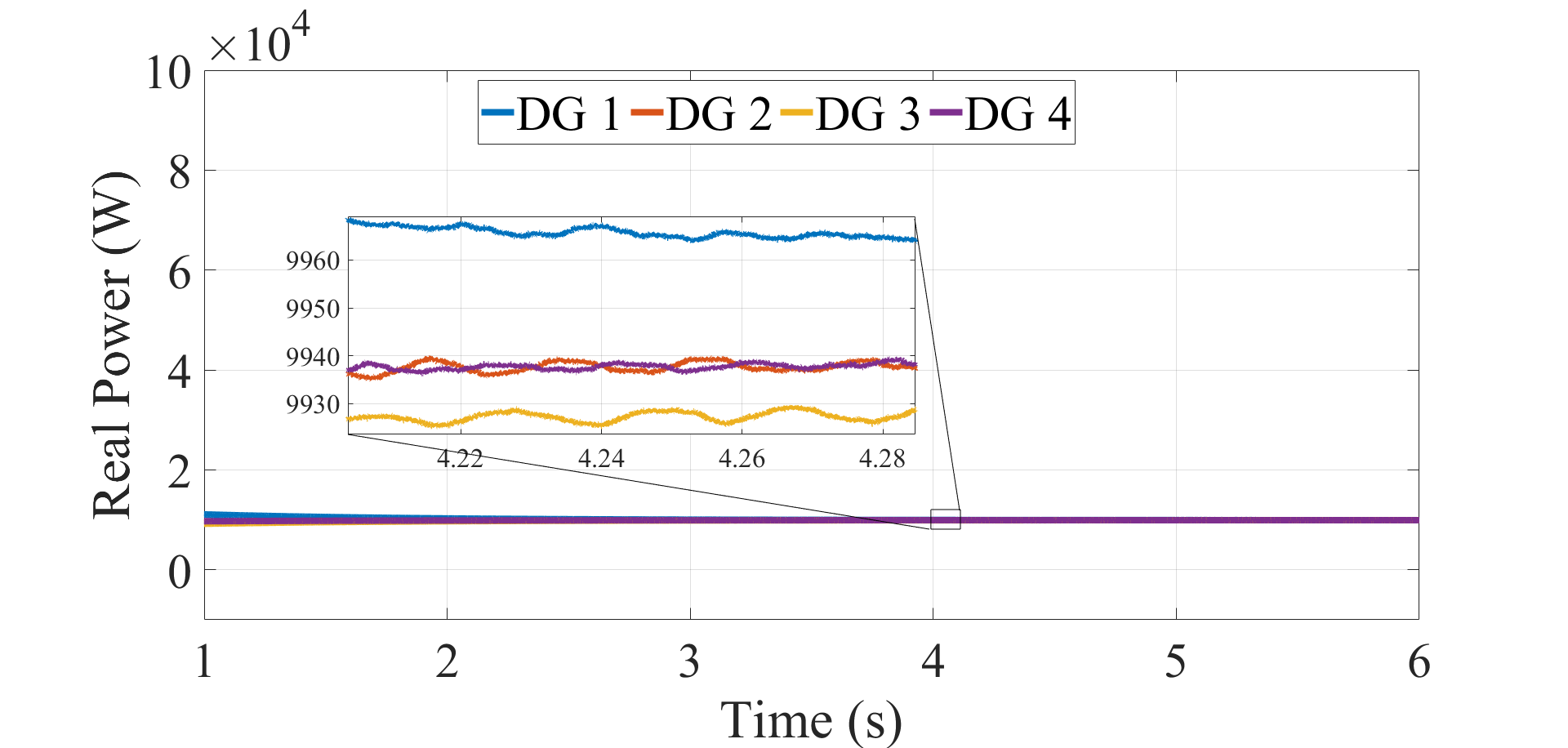}
  \vspace{-4mm}
  \caption{Load sharing among DGs during nominal operation.}
  \label{fig:6}
\end{figure}

\begin{figure}[t]
\centering
  \includegraphics[width=0.9\linewidth]{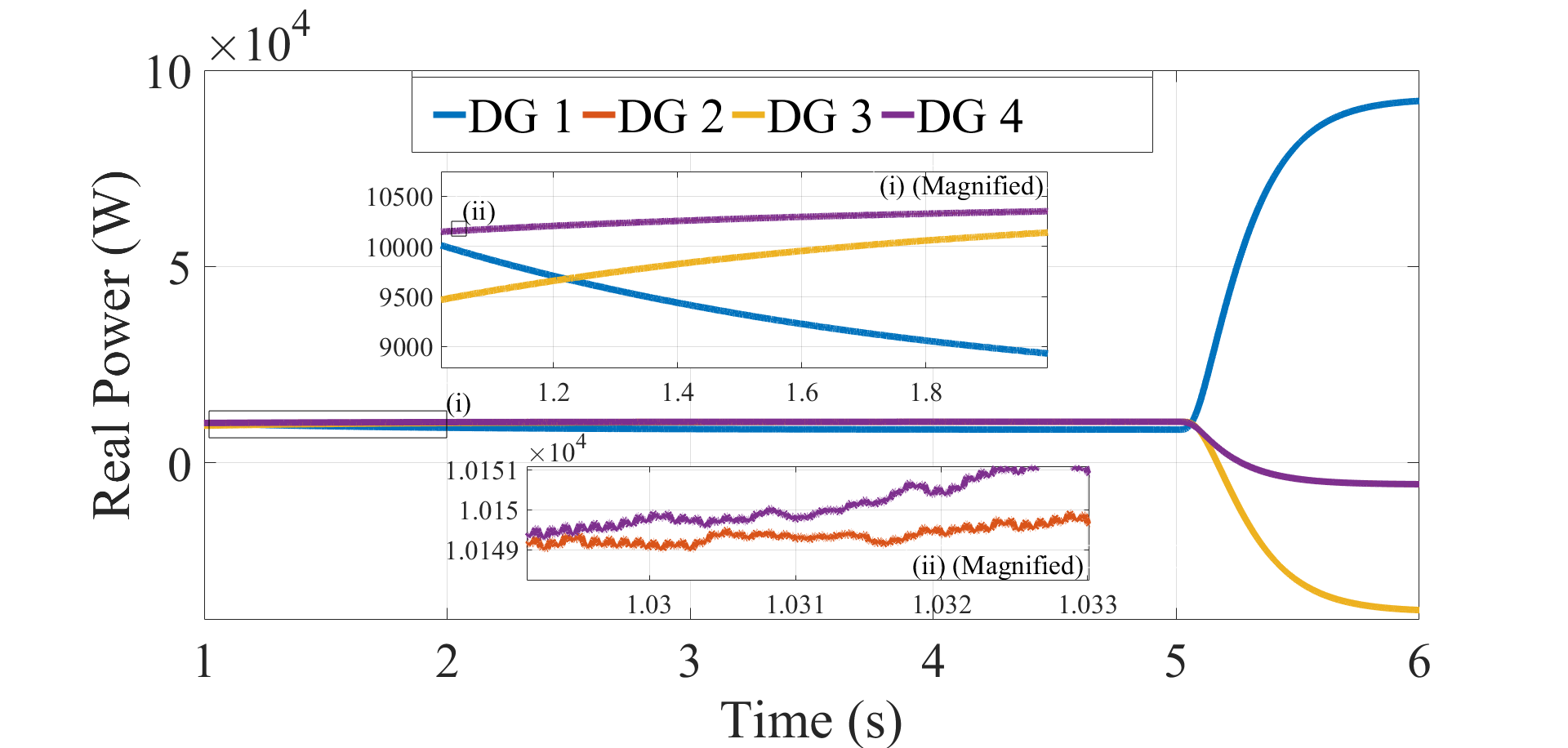}
  \vspace{-4mm}
  \caption{Impact of rootkit manipulations on the load sharing scheme.}
  \label{fig:7}
\end{figure}






\subsection{Target: Disturbance in Load Sharing}
\textcolor{black}{
In the attack-free scenario, all DGs in the simulated system share the total load demand equally (load sharing pattern shown in Fig. \ref{fig:6}). However, during the attack phase, the rootkit disrupts the existing load sharing pattern. After the vulnerable agent identification, the malware establishes a time frame ($t=5s$) within which manipulations will be introduced. For instance, an attack could be initiated during peak demand when the DGs operate close to their generation limits. Fig. \ref{fig:7} shows the disruption in load sharing caused by the rootkit through controller-level manipulation. Continued manipulations destabilize the system (after $t=5s$) creating instabilities.}
\vspace{-4mm}
\section{Conclusions and Future Work}
\vspace{-1mm}
\textcolor{black}{
The impact of process-aware rootkit attacks represents a severe threat for power systems. The presented rootkit can learn the MG's operational patterns. We demonstrate how such information can help mount crippling attacks against the local DGs and the PCC level. Rootkits can achieve persistence by masking their presence. Our future work will focus on developing detection and mitigation mechanisms, e.g.,  moving target defences, stable kernel represenations \cite{zografopoulos2021detection}, etc. thwarting the adversarial learning behavior of rootkits. }


\vspace{-2mm}
\bibliographystyle{IEEEtran}
\bibliography{biblio}

\begin{thebibliography}{10}
\providecommand{\url}[1]{#1}
\csname url@samestyle\endcsname
\providecommand{\newblock}{\relax}
\providecommand{\bibinfo}[2]{#2}
\providecommand{\BIBentrySTDinterwordspacing}{\spaceskip=0pt\relax}
\providecommand{\BIBentryALTinterwordstretchfactor}{4}
\providecommand{\BIBentryALTinterwordspacing}{\spaceskip=\fontdimen2\font plus
\BIBentryALTinterwordstretchfactor\fontdimen3\font minus
  \fontdimen4\font\relax}
\providecommand{\BIBforeignlanguage}[2]{{%
\expandafter\ifx\csname l@#1\endcsname\relax
\typeout{** WARNING: IEEEtran.bst: No hyphenation pattern has been}%
\typeout{** loaded for the language `#1'. Using the pattern for}%
\typeout{** the default language instead.}%
\else
\language=\csname l@#1\endcsname
\fi
#2}}
\providecommand{\BIBdecl}{\relax}
\BIBdecl

\bibitem{rath2020cyber}
S.~Rath \emph{et~al.}, ``A cyber-secure distributed control architecture for
  autonomous ac microgrid,'' \emph{IEEE Systems Journal}, vol.~15, no.~3, 2021.

\bibitem{konstantinou2017gps}
C.~Konstantinou \emph{et~al.}, ``Gps spoofing effect on phase angle monitoring
  and control in a real-time digital simulator-based hardware-in-the-loop
  environment,'' \emph{IET Cyber-Physical Systems: Theory \& Applications},
  vol.~2, no.~4, pp. 180--187, 2017.

\bibitem{zografopoulos2021cyber}
I.~Zografopoulos \emph{et~al.}, ``Cyber-physical energy systems security:
  Threat modeling, risk assessment, resources, metrics, and case studies,''
  \emph{IEEE Access}, vol.~9, pp. 29\,775--29\,818, 2021.

\bibitem{powers2015whitelist}
J.~Powers \emph{et~al.}, ``Whitelist malware defense for embedded control
  system devices,'' in \emph{Saudi Arabia Smart Grid (SASG)}.\hskip 1em plus
  0.5em minus 0.4em\relax IEEE, 2015.

\bibitem{8988665}
P.~Krishnamurthy \emph{et~al.}, ``Stealthy rootkits in smart grid
  controllers,'' in \emph{International Conference on Computer Design (ICCD)},
  2019, pp. 20--28.

\bibitem{xenofontos2021consumer}
C.~Xenofontos \emph{et~al.}, ``Consumer, commercial and industrial iot(in)
  security: attack taxonomy and case studies,'' \emph{IEEE Internet of Things
  Journal}, vol.~9, no.~1, pp. 199--221, 2022.

\bibitem{reeves2012intrusion}
J.~Reeves \emph{et~al.}, ``Intrusion detection for resource-constrained
  embedded control systems in the power grid,'' \emph{International Journal of
  Critical Infrastructure Protection}, vol.~5, no.~2, pp. 74--83, 2012.

\bibitem{10.1145/3447555.3466576}
S.~Rath, I.~Zografopoulos, and C.~Konstantinou, ``Stealthy rootkit attacks on
  cyber-physical microgrids: Poster,'' in \emph{Proceedings of the 12th ACM
  Int'l Conference on Future Energy Systems}.\hskip 1em plus 0.5em minus
  0.4em\relax ACM, 2021, p. 294–295.

\bibitem{MITRE}
\BIBentryALTinterwordspacing
MITRE. {Rootkit}. [Online]. Available: \url{https://tinyurl.com/2p83p89n}
\BIBentrySTDinterwordspacing

\bibitem{kuruvila2020hardware}
A.~P. Kuruvila \emph{et~al.}, ``Hardware-assisted detection of firmware attacks
  in inverter-based cyberphysical microgrids,'' \emph{International Journal of
  Electrical Power \& Energy Systems}, vol. 132, p. 107150, 2021.

\bibitem{9595243}
M.~Leng, S.~Sahoo, and F.~Blaabjerg, ``Stability investigation of dc microgrids
  under stealth cyber attacks,'' in \emph{2021 IEEE Energy Conversion Congress
  and Exposition (ECCE)}, 2021, pp. 1427--1432.

\bibitem{lakshminarayana2022loadaltering}
S.~Lakshminarayana, J.~Ospina, and C.~Konstantinou, ``Load-altering attacks
  under covid-19 low-inertia conditions,'' 2022.

\bibitem{bidram2017cooperative}
A.~Bidram \emph{et~al.}, \emph{Cooperative synchronization in distributed
  microgrid control}.\hskip 1em plus 0.5em minus 0.4em\relax Springer, 2017.

\bibitem{mnatsakanyan2014novel}
A.~Mnatsakanyan and S.~W. Kennedy, ``A novel demand response model with an
  application for a virtual power plant,'' \emph{IEEE Transactions on Smart
  Grid}, vol.~6, no.~1, pp. 230--237, 2014.

\bibitem{baringo2018day}
A.~Baringo, L.~Baringo, and J.~M. Arroyo, ``Day-ahead self-scheduling of a
  virtual power plant in energy and reserve electricity markets under
  uncertainty,'' \emph{IEEE Transactions on Power Systems}, vol.~34, no.~3,
  2018.

\bibitem{keliris2019open}
A.~Keliris \emph{et~al.}, ``Open source intelligence for energy sector
  cyberattacks,'' in \emph{Critical infrastructure security and
  resilience}.\hskip 1em plus 0.5em minus 0.4em\relax Springer, 2019.

\bibitem{wang2016malicious}
X.~Wang \emph{et~al.}, ``Malicious firmware detection with hardware performance
  counters,'' \emph{IEEE Transactions on Multi-Scale Computing Systems},
  vol.~2, no.~3, pp. 160--173, 2016.

\bibitem{anubi2019enhanced}
O.~M. Anubi and C.~Konstantinou, ``Enhanced resilient state estimation using
  data-driven auxiliary models,'' \emph{IEEE Transactions on Industrial
  Informatics}, vol.~16, no.~1, pp. 639--647, 2019.

\bibitem{habtom1997estimation}
R.~Habtom and L.~Litz, ``Estimation of unmeasured inputs using recurrent neural
  networks and the extended kalman filter,'' in \emph{Proceedings of
  International Conference on Neural Networks}, vol.~4.\hskip 1em plus 0.5em
  minus 0.4em\relax IEEE, 1997.

\bibitem{sieberg2021hybrid}
P.~M. Sieberg \emph{et~al.}, ``Hybrid state estimation-a contribution towards
  reliability enhancement of artificial neural network estimators,'' \emph{IEEE
  Transactions on Intelligent Transportation Systems}, 2021.

\bibitem{zhan2006neural}
R.~Zhan and J.~Wan, ``Neural network-aided adaptive unscented kalman filter for
  nonlinear state estimation,'' \emph{IEEE Signal Processing Letters}, vol.~13,
  no.~7, pp. 445--448, 2006.

\bibitem{zografopoulos2021detection}
I.~Zografopoulos and C.~Konstantinou, ``Detection of malicious attacks in
  autonomous cyber-physical inverter-based microgrids,'' \emph{IEEE
  Transactions on Industrial Informatics}, 2021.

\end{thebibliography}

\end{document}